\documentclass[twocolumn]{aastex631}

\shorttitle{Improvement of the PS1 Photometric Calibration with LAMOST and Gaia}
\shortauthors{Xiao et al.}
\graphicspath{{./}{figures/}}

\usepackage{subfigure}
\usepackage{amsmath}
\newcommand{\grizy}{$g$, $r$, $i$, $z$, and $y$}
\newcommand{\bprp}{$G_{\rm BP}-G_{\rm RP}$}
\newcommand{\bprpo}{$(G_{\rm BP}-G_{\rm RP})_{\rm 0}$}
\newcommand{\bpg}{$G_{\rm BP}-g$}
\newcommand{\rpr}{$G_{\rm RP}-r$}
\newcommand{\rpi}{$G_{\rm RP}-i$}
\newcommand{\rpz}{$G_{\rm RP}-z$}
\newcommand{\rpy}{$G_{\rm RP}-y$}
\newcommand{\feh}{$\rm [Fe/H]$}

\newcommand{\ebv}{$E(B-V)_{\rm SFD}$}
\newcommand{\ebprp}{$E(G_{\rm BP}-G_{\rm RP})$}

\begin{document}

\title{Improvement of the Pan-STARRS Photometric Calibration with LAMOST and Gaia}

\author{
Kai Xiao\altaffilmark{1,2}
Haibo Yuan\altaffilmark{1,2}
Bowen Huang\altaffilmark{1,2}
Ruoyi Zhang\altaffilmark{1,2}
Lin Yang\altaffilmark{3}
Shuai Xu\altaffilmark{1,2}
}
\altaffiltext{1}{Institute for Frontiers in Astronomy and Astrophysics, Beijing Normal University, Beijing, 102206, China; email: yuanhb@bnu.edu.cn}
\altaffiltext{2}{Department of Astronomy, Beijing Normal University, Beijing, 100875, People's Republic of China}
\altaffiltext{3}{College of Artificial Intelligence, Beijing Normal University No.19, Xinjiekouwai St, Haidian District, Beijing, 100875, P.R.China}

\journalinfo{submitted to ApJS}
\submitted{Received 2023 June 7; accepted 2023 August 8}

\begin{abstract}
In this work, we perform the re-calibration of PS1 photometry by correcting for position-dependent systematic errors using the spectroscopy-based Stellar Color Regression method (SCR), the photometry-based SCR method (SCR$'$), and the Gaia XP synthetic photometry method (XPSP). We confirm the significant large-scale and small-scale spatial variation of magnitude offsets for all the $grizy$ filters.
We show that the PS1 photometric calibration precisions in the $grizy$ filters are around 5--7\,mmag when averaged over 14$'$ regions. We note a much larger calibration error up to 0.04 mag in the Galactic plane, which is probably caused by the systematic errors of the PS1 magnitudes in crowded fields. The results of the three methods are consistent with each other within 1--2\,mmag or better for all the filters. We provide two-dimensional maps and a python package ({\url{https://doi.org/10.12149/101283}}) to correct for position-dependent magnitude offsets of PS1, which can be used for high-precision investigations and as a reference to calibrate other surveys.
\end{abstract}


\section{Introduction} \label{sec:intro}

Pan-STARRS1 (PS1; \citealt{2012ApJ...750...99T}), as the first part of the Pan-STARRS (\citealt{2002SPIE.4836..154K, 2010SPIE.7733E..0EK}), has imaged three quarters of the sky of north of $\rm decl.=-30^{\circ}$ repeatly in the \grizy~filters.
As one of the best calibrated photometric surveys (\citealt{2020ApJS..251....6M}), it has been widely used as the reference to calibrate other surveys (e.g., \citealt{2015ApJ...815..117S,2016ApJ...822...66F,2017AJ....153..276Z,2018PASP..130h5001Z,2019A&A...631A.119L, 2021A&A...654A..61L}).

To validate and improve the PS1 photometric calibration, \citet{2022AJ....163..185X} performed an independent test using the spectroscopy-based Stellar Color Regression (SCR) method. The SCR method was proposed by \cite{2015ApJ...799..133Y} and has since been widely used (e.g., \citealt{2021ApJ...907...68H, 2021ApJ...908L..24Y, 2021ApJ...909...48N, 2021ApJ...908L..14N,2022ApJS..259...26H}; see also the review by \citealt{2022SSPMA..52B9503H}). \citet{2022AJ....163..185X} selected approximately 1.5 million LAMOST-Gaia-PS1 FGK dwarf stars as standards, using spectroscopic data from the Large Sky Area Multi-Object Fiber Spectroscopic Telescope Data Release 7 (LAMOST DR7; \citealt{2015RAA....15.1095L}) and photometric data from the corrected Gaia Early Data Release 3 (EDR3; \citealt{2021A&A...649A...1G,2021A&A...650C...3G}). The results showed that the PS1 calibration precision is around $4\sim 5$ millimagnitudes when averaged over $20'$ regions in the $grizy$ filters, consistent with an internal precision of better than $1\%$\,mag \citep{2012ApJ...756..158S}. However, significant large-scale and small-scale spatial variations of the magnitude offsets, i.e., the calibration errors, were found. Moderate magnitude-dependent errors were also found and corrected for the $grizy$ filters. The two-dimensional correction maps have been provided but are restricted to the LAMOST footprint.

Recently, \citet{2022ApJS..258...44X} delivered reliable photometric metallicities for approximately 27 million stars with $|b|>10^{\circ}$, $10<G<16$, and $E(B-V)_{\rm SFD}\le0.5$. Here $E(B-V)_{\rm SFD}$ refers to reddening values from the 2D reddening map of \citet{1998ApJ...500..525S}. They achieved this by combining spectroscopic data from the LAMOST DR7 and photometric data from the corrected Gaia EDR3 via the metallicity-dependent stellar locus. This sample can serve as new standard stars for recalibrating the PS1 data via a photometry-based SCR method.

More recently, the Gaia Data Release 3 \citep{2021A&A...652A..86C} provided calibrated low-resolution spectra, $\lambda/\delta \lambda \sim 50$, from the BP and RP (hereafter shortened as XP) spectrophotometers for about 220 million sources. Based on the XP spectra, \citet{2022arXiv220606215G} also provided a Gaia Synthetic Photometry Catalog (GSPC).
The GSPC catalog contains standardized photometric data in several widely-used bands. Additionally, a Python software tool called $\texttt{GaiaXPy}$\footnote{\url{https://gaia dpci.github.io/GaiaXPy-website/}} is provided to generate synthetic standardized photometry for a number of photometric systems, such as PS1. As pointed out in \citet{2022arXiv220606215G}, XP Synthetic Photometry (XPSP, hereafter) provides a great opportunity for the validation and/or re-calibration of existing or future photometric surveys.

In this work, we take advantage of the aforementioned new data-sets available to perform the re-calibration of the PS1 photometry for the whole survey footprint. The paper is organized as follows: Section\,\ref{sec:data} and Section\,\ref{sec:method} introduce the data and methods used in this work. The results are presented in Section\,\ref{sec:res} and discussed in Section\,\ref{sec:discussion}. Finally, Section\,\ref{sec:conclusion} provides our conclusions.

\section{Data} \label{sec:data}
\subsection{Pan-STARRS 1 Data Release 1} \label{sec:ps1}
The PS1 survey has imaged three-quarters of the sky using five broadband filters ($g$, $r$, $i$, $z$, $y$). Its first public data release (DR1) includes the results of the third full reduction of the Pan-STARRS $3\pi$ Survey (\citealt{2020ApJS..251....6M}).

\subsection{Gaia Early Data Release 3} \label{sec:gaia}
The Gaia EDR3 (\citealt{2021A&A...649A...1G,2021A&A...650C...3G}) has provided the best photometric data for approximately 1.8 billion stars, with uniform calibration at the mmag level in the $G$, $G_{\rm BP}$, and $G_{\rm RP}$ bands. 
\citet{2021ApJ...908L..24Y} found and corrected for magnitude-dependent systematic errors of around 1\%\,mag for the $G$, $G_{\rm BP}$, and $G_{\rm RP}$ bands of Gaia EDR3, using approximately 10,000 Landolt standard stars. 
By comparing the results of \cite{2021ApJ...908L..24Y} and \cite{2021ApJ...908L..14N}, it was observed that the Gaia colors obtained from completely different data and methods exhibit a remarkable consistency within 1\,mmag. This outstanding consistency is further supported by \cite{2021ApJS..255...20A} and \cite{2021ApJ...922..211N}, highlighting the reliability of Gaia photometric data. These findings pave the way for the optimal utilization of Gaia EDR3 photometry in high-precision investigations and as a reference for calibrating other surveys.
This paper uses the magnitudes of $G$, $G_{\rm BP}$, and $G_{\rm RP}$ as corrected by \citet{2021ApJ...908L..24Y} by default.

\subsection{Gaia Data Release 3} \label{sec:gaia}
The Gaia DR3 (\citealt{2021A&A...652A..86C}) has provided low resolution ($\lambda/\Delta \lambda\sim$ 50) spectra from the XP spectro-photometers for around 220 million sources, with most having $G < 17.65$. The XP spectra cover wavelengths from 330 to 1050\,nm and have been precisely calibrated both internally \citep{2021A&A...652A..86C,2022arXiv220606143D} and externally \citep{2022arXiv220606205M}.

\begin{figure}[ht!]
\resizebox{\hsize}{!}{\includegraphics{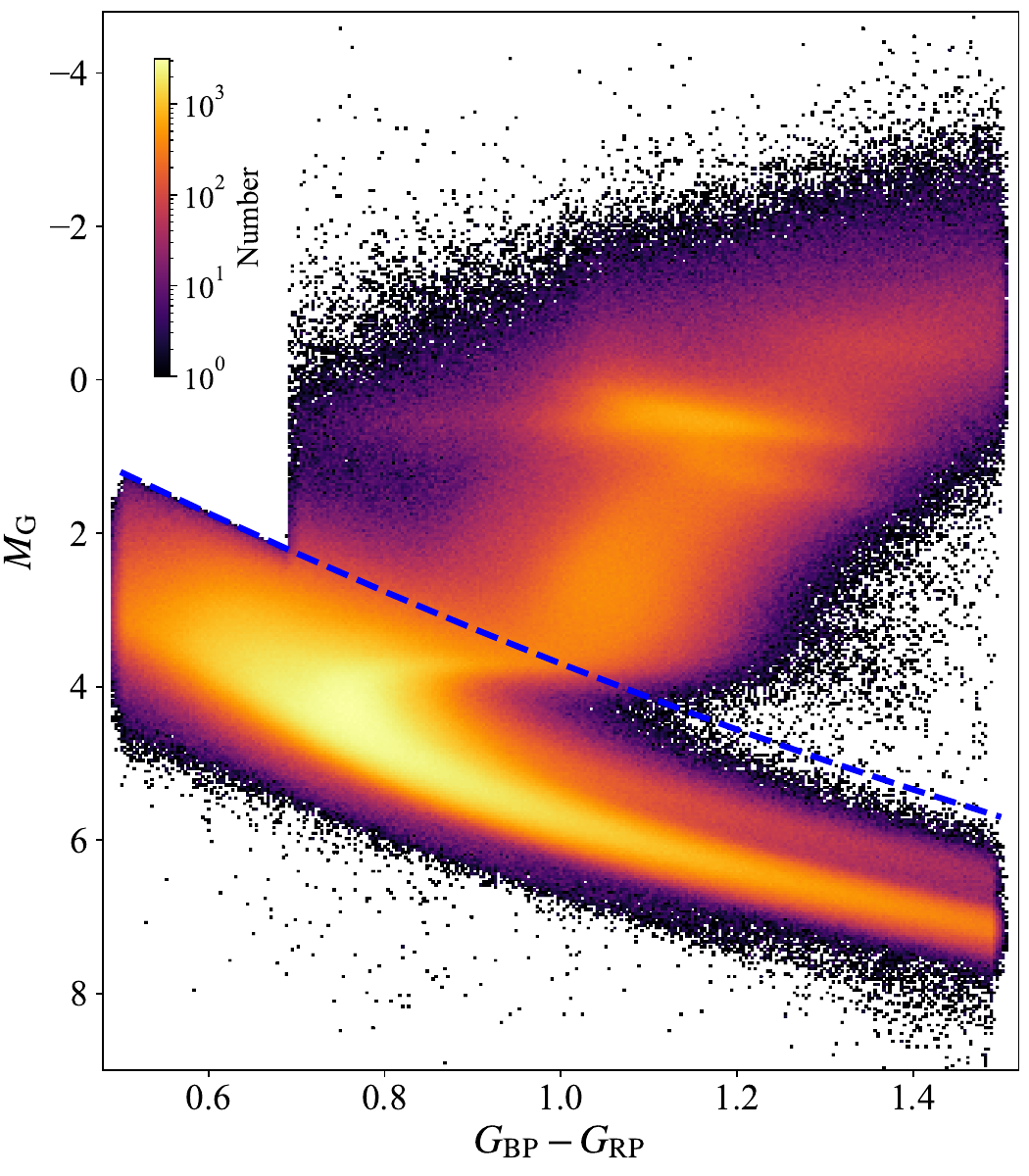}}
\caption{{\small HR diagram for stars in the $g$ band after combining the PS1 data with the catalog of 
\citet{2022ApJS..258...44X}. 
The color bar in the upper-left corner indicates the stellar number densities. The blue dotted line at $m_{\rm G} = -(G_{\rm BP}-G_{\rm RP})^2 + 6.5\times(G_{\rm BP}-G_{\rm RP})-1.8$ given by \citet{2022ApJS..258...44X} marks the boundary between the dwarfs and the giants. 
\cite{2022ApJS..258...44X} applied the color cut of $0.5<(G_{\rm BP}-G_{\rm RP})<1.5$ for dwarfs and $0.7<(G_{\rm BP}-G_{\rm RP})<1.4$ for giants during the photometric metallicities construction.
}}
\label{Fig:mcd}
\end{figure}

\begin{deluxetable}{cccccc}[ht!]
\tablecaption{The coefficients used to obtain $\delta \bf R$ as a function of $E(B-V)_{\rm SFD}$ in the five colors. $\delta R_{\rm color}=a_{\rm 0}\cdot x^4+a_{\rm 1}\cdot x^3+a_{\rm 2}\cdot x^2+a_{\rm 3}\cdot x+a_{\rm 4}$, where $x$ is $E(B-V)_{\rm SFD}$.
\label{tab:1}}
\tablehead{
\colhead{Color} & \colhead{$a_{\rm 0}$} & \colhead{$a_{\rm 1}$} & \colhead{$a_{\rm 2}$} & \colhead{$a_{\rm 3}$} & \colhead{$a_{\rm 4}$}}
\startdata
$G_{\rm BP}-g$ & $+$0.020 & $-$0.348 & $+$1.976 & $-$4.693 & $+$1.938 \\
$G_{\rm RP}-r$ & $+$0.001 & $+$0.070 & $-$0.479 & $+$1.173 & $-$0.488 \\
$G_{\rm RP}-i$ & $+$0.009 & $-$0.122 & $+$0.605 & $-$1.288 & $+$0.497 \\
$G_{\rm RP}-z$ & $+$0.013 & $-$0.293 & $+$1.671 & $-$3.761 & $+$1.476 \\
$G_{\rm RP}-y$ & $+$0.012 & $-$0.275 & $+$1.552 & $-$3.482 & $+$1.367
\enddata
\end{deluxetable}

\section{obtain the magnitude offsets $\Delta {\bf m}(\rm R.A., decl.)$} 
\label{sec:method}
In this section, we introduce two methods used to obtain the magnitude offsets $\Delta {\bf m}(\rm R.A., decl.)$ of the PS1 data. Note that the PS1 DR1 magnitudes ${\bf m_{\rm PS1}}$ mentioned here refer to the PSF magnitudes after correcting for magnitude-dependent errors as described in \citet{2022AJ....163..185X}.
The correction of magnitude-dependent systematic errors in PS1 using the SCR method does not introduce spatial-dependent systematic errors for PS1.

\begin{figure*}[ht!] \centering
\resizebox{\hsize}{!}{\includegraphics{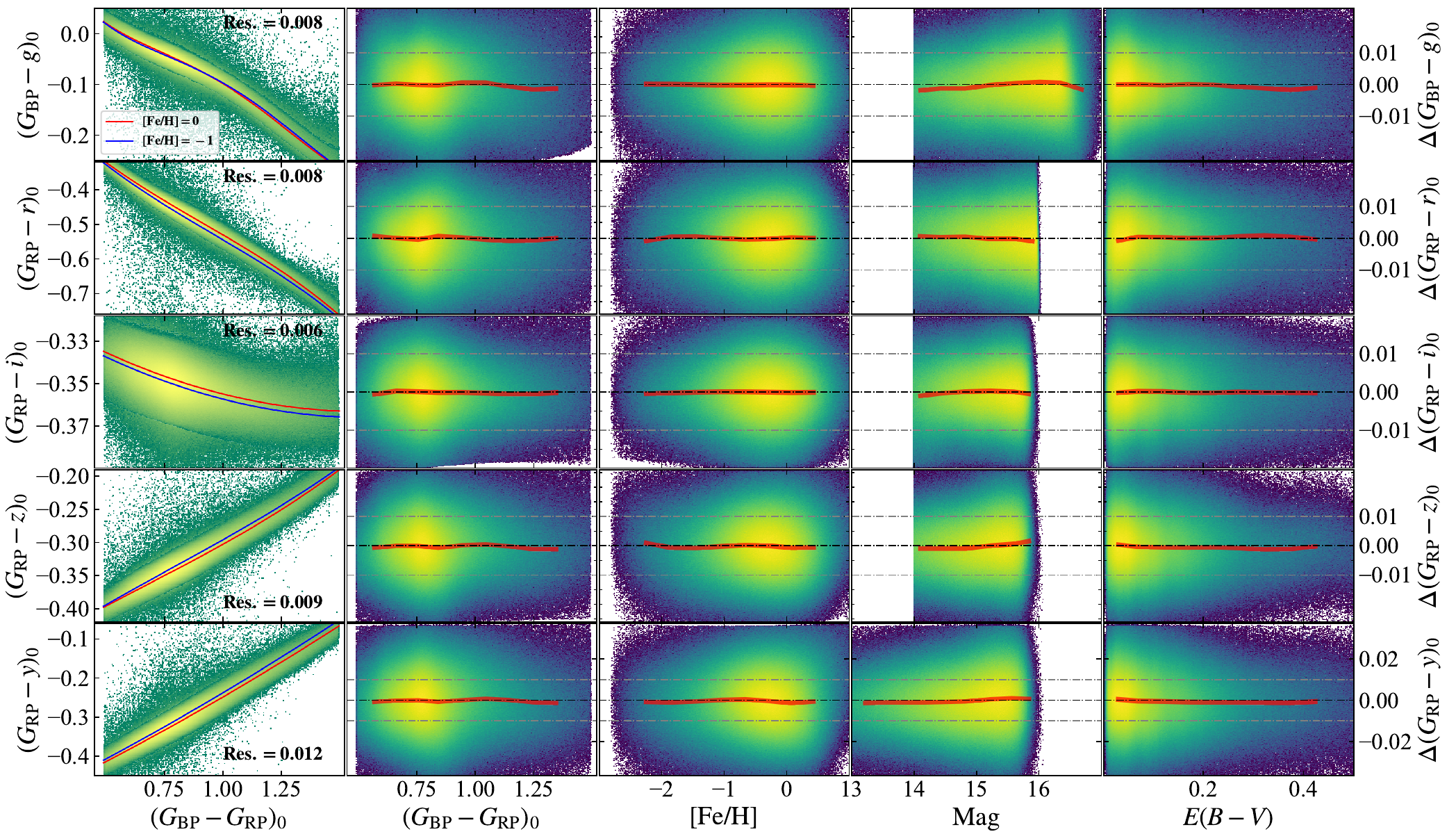}}
\caption{{\small  Two-dimensional polynomial fitting of the intrinsic colors as functions of \bprpo~and \feh~for the SCR$'$ calibration stars. From top to bottom are for the \bpg, \rpr, \rpi, \rpz,~and \rpy~colors, respectively. The left column shows the fitting results after $3\sigma$ clipping. The red and blue curves represent results for $\rm [Fe/H]$ = 0 and $-$1, respectively, and the standard deviations of the fitting residuals are labeled. The 2nd, 3rd, 4th and 5th columns plot residuals against \bprpo, \feh, magnitude, and \ebv, respectively, with the medians over-plotted in red.
}}
\label{Fig:intrinsic_colors_fitting}
\end{figure*}

\begin{deluxetable*}{ccccccccccc}[ht!]
\tablecaption{The coefficients used to obtain intrinsic colors as functions of $(G_{\rm BP}-G_{\rm RP})_{\rm 0}$ and \feh. 
$C^{\rm mod}_{\rm 0}={a_0}\cdot x^3+{a_1}\cdot y^3+{a_2}\cdot x^2\cdot y+{a_3}\cdot x\cdot y+{a_4}\cdot x^2+{a_5}\cdot y^2+{a_6}\cdot x\cdot y+{a_7}\cdot x+{a_8}\cdot y+{a_9}$, where $x$ is $(G_{\rm BP}-G_{\rm RP})_{\rm 0}$, and $y$ is ${\rm [Fe/H]}$. \label{tab:2}}
\tablehead{
\colhead{Intrinsic Color} & \colhead{$a_0$} & \colhead{$a_1$} & \colhead{$a_2$} & \colhead{$a_3$} & \colhead{$a_4$} & \colhead{$a_5$} & \colhead{$a_6$} & \colhead{$a_7$} & \colhead{$a_8$} & \colhead{$a_9$}}
\startdata
$(G_{\rm BP}-g)_{\rm 0}$ & $-$1.8181 & $+$0.0008 & $-$0.2147 & $-$0.0111 & $+$2.5337 & $+$0.0075 & $+$0.1805 & $-$1.7029 & $-$0.0426 & $+$0.4357 \\
$(G_{\rm RP}-r)_{\rm 0}$ & $-$0.1852 & $-$0.0009 & $-$0.0262 & $+$0.0055 & $+$0.5058 & $-$0.0031 & $+$0.0617 & $-$0.8501 & $-$0.0184 & $-$0.0008 \\
$(G_{\rm RP}-i)_{\rm 0}$ & $-$       & $-$       & $-$       & $-$       & $+$0.0228 & $+$0.0007 & $+$0.0011 & $-$0.0721 & $+$0.0021 & $-$0.3055 \\
$(G_{\rm RP}-z)_{\rm 0}$ & $+$0.0452 & $+$0.0013 & $+$0.0148 & $-$0.0042 & $-$0.0954 & $+$0.0054 & $-$0.0358 & $+$0.2513 & $+$0.0142 & $-$0.5035 \\
$(G_{\rm RP}-y)_{\rm 0}$ & $-$       & $-$       & $-$       & $-$       & $+$0.0350 & $-$0.0029 & $-$0.0088 & $+$0.2772 & $-$0.0050 & $-$0.5588
\enddata
\end{deluxetable*}

\begin{figure*}[htbp]
\resizebox{\hsize}{!}{\includegraphics{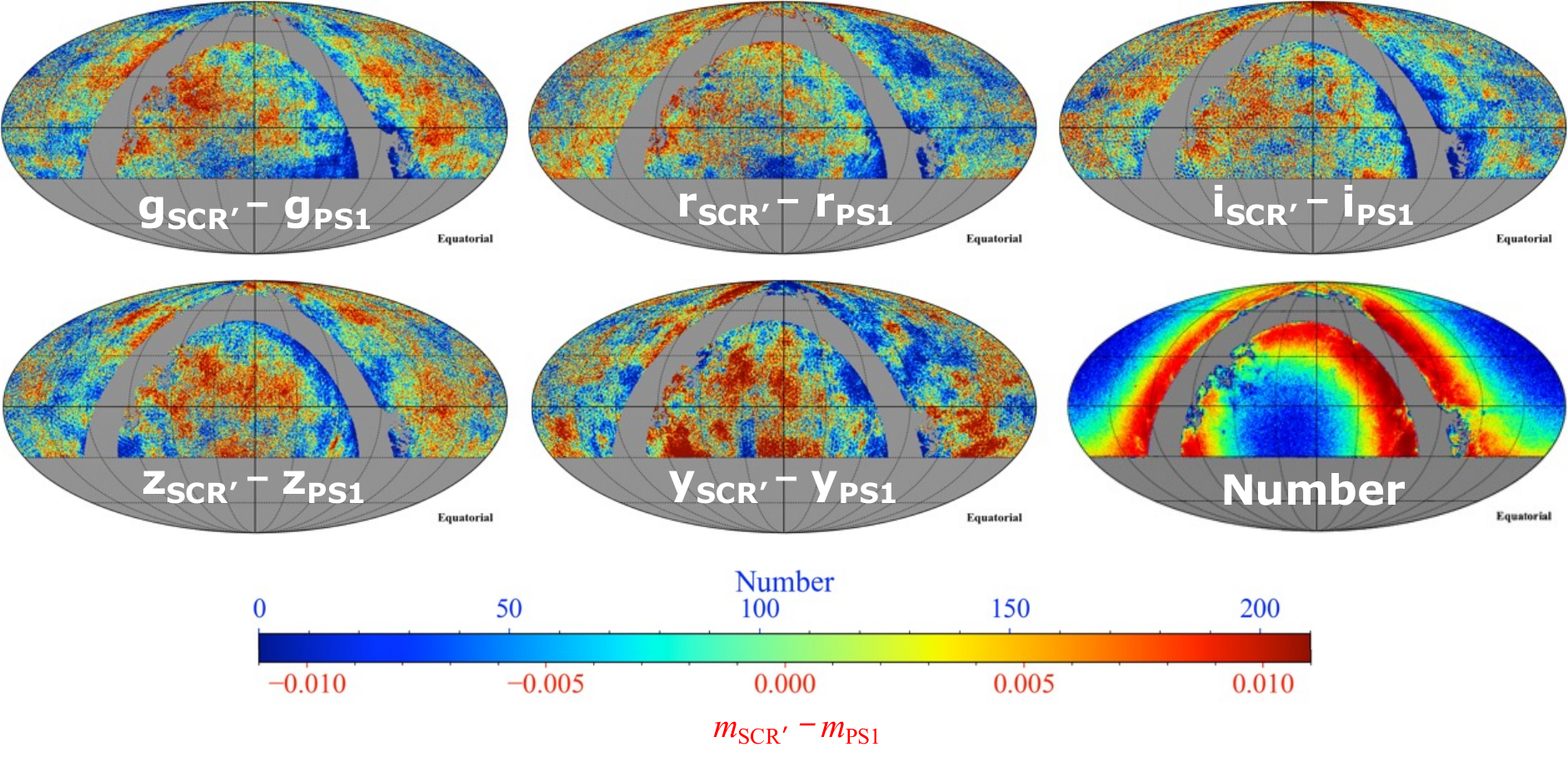}}
\caption{{\small Spatial variations of magnitude offsets obtained by the SCR$'$ method in the $g$, $r$, $i$, $z$ and $y$ bands, and spatial distribution of the calibration sample stars for the $g$ band in the Equatorial coordinate system. Color bars are overplotted to the bottom. HEALPix index is level 8, corresponding to a physical scale of 14$'$.}}\label{Fig:newscr}
\end{figure*}

\begin{figure*}[htbp]
\resizebox{\hsize}{!}{\includegraphics{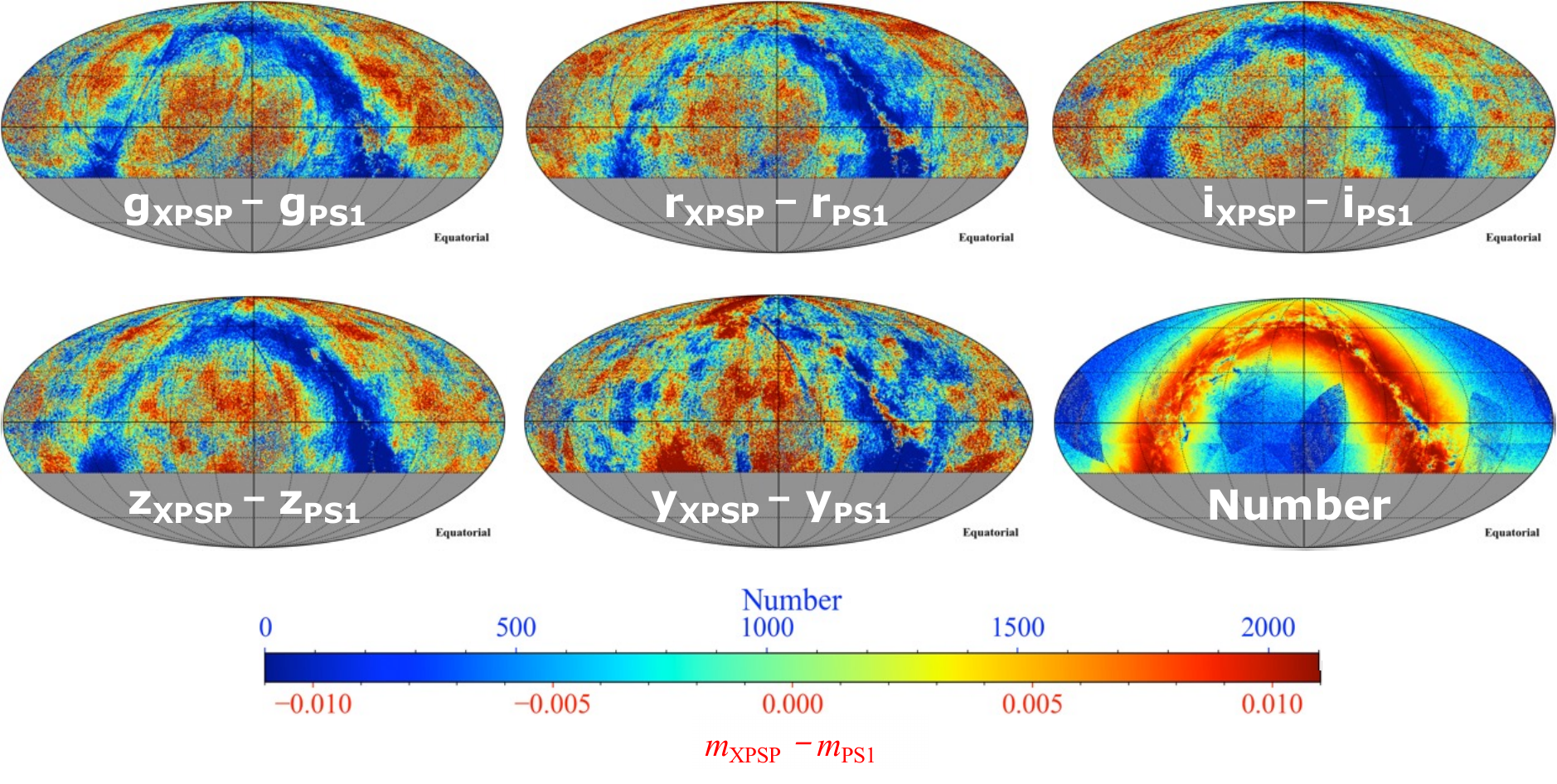}}
\caption{{\small Same as Figure\,\ref{Fig:newscr}, but for the XPSP method.}}\label{Fig:gspc}
\end{figure*}

\subsection{The photometry-based SCR (SCR$'$) method} \label{sec:m2}
As noted by \citet{2022ApJS..259...26H} and \citet{2022SSPMA..52B9503H}, the key idea of the SCR method is that the intrinsic colors of stars are relatively simple, and can be fully determined by a small number of stellar parameters. Different forms of stellar parameters may be used in different situations. In \citet{2022AJ....163..185X}, the spectroscopy-based SCR method was used, where the intrinsic colors of dwarf stars are functions of $T_{\rm eff}$ and $\rm [Fe/H]$. However, to take advantage of the much larger sample with precise photometric metallicities from \citet{2022ApJS..258...44X}, we adopt the photometry-based SCR method using the metallicity-dependent stellar locus (\citealt{2015ApJ...799..134Y}), where the intrinsic colors of stars are predicted using the \bprpo~color and metallicity \feh.
As a result, we construct approximately four times as many standard stars as in \citet{2022AJ....163..185X}, covering almost the entire PS1 footprint, except for the disk regions. To distinguish the results from the spectroscopy-based SCR method, we denote the results obtained by the photometry-based SCR method with the subscript $\rm SCR'$, which has also been employed in the photometric calibration of SAGES Nanshan One-meter Wide-field Telescope $gri$ imaging data (Xiao et al. 2023).
The detailed description of the method is as follows.

We combine the PS1 DR1 photometric data with the catalog from \citet{2022ApJS..258...44X}, using a cross-matching radius of $1''$. We select dwarf stars (see Figure\,\ref{Fig:mcd}) as our calibration samples, subject to the following constraints:
\begin{enumerate}
  \item[a.] mag\{$g, r, i, z$\} $>$ 14 and mag\{$y$\} $>$ 13 to avoid saturation;
  \item[b.] error\{$g, r, i, z, y$\} $<$ 0.02\,mag;
  \item[c.] \texttt{phot}\_\texttt{bp}\_\texttt{rp}\_\texttt{excess}\_\texttt{factor} $<$ $1.3+0.06\times(G_{\rm BP}-G_{\rm RP})^2$ to avoid bad Gaia photometry;
\item[d.] the vertical distance to the Galactic plane is larger than 300\,pc to avoid invalid reddening values from the dust reddening map of \citet{1998ApJ...500..525S}.
\end{enumerate}
Finally, we select 6,686,051, 6,333,912, 6,126,640, 6,096,912, and 6,991,511 calibration stars in the $g$, $r$, $i$, $z$, and $y$ bands, respectively.

Five colors ${\bf C}=(G_{\rm BP}-g, G_{\rm RP}-r, G_{\rm RP}-i, G_{\rm RP}-z, G_{\rm RP}-y)^{\rm T}$ are adopted for the $grizy$ filters of the PS1, as in \citet{2022AJ....163..185X}. The intrinsic colors of stars (denoted by ${\bf C_{\rm 0}}$) can be estimated as ${\bf C_{\rm 0}}={\bf C} -{\bf R} \times E(B-V)$, where $R$ refers to reddening coefficient, or obtained from the metallicity-dependent stellar loci. The latter are derived by fitting third-order two-dimensional polynomials (with 10 free parameters) to the $grz$ bands and second-order polynomials (with 6 free parameters) to other bands, taking into account the dependence of stellar colors on metallicity.
 
The reddening correction is performed using the dust reddening map from \citet{1998ApJ...500..525S} and empirical temperature- and reddening-dependent reddening coefficients from \citet{2023ApJS..264...14Z}. The estimation of temperatures for all calibration stars is provided in the Appendix. However, a slight dependence (of about 0.01 mag or smaller) on $E(B-V)_{\rm SFD}$ is observed in the residuals when fitting intrinsic colors as a function of $(G_{\rm BP}-G_{\rm RP})_{\rm 0}$ and metallicity $\rm [Fe/H]$ of calibration stars. This dependence is probably due to the fact that   \citet{2023ApJS..264...14Z} used only a second-order binary function to describe the dependence of reddening coefficients on temperature and reddening to avoid over-fitting. Therefore, we use a fourth-order polynomial as a function of $E(B-V)_{\rm SFD}$ to fit the residuals divided by $E(B-V)_{\rm SFD}$. The corresponding fitting parameters are listed in Table\,\ref{tab:1}. The corrected reddening coefficients ${\bf R'}$ can be obtained as ${\bf R}+{\bf \delta R}(E(B-V)_{\rm SFD})$. Iterations are performed.

The final fitting results of the intrinsic colors as a function of \bprpo~and \feh~for the SCR$'$ method are presented in Figure\,\ref{Fig:intrinsic_colors_fitting}, and the corresponding fitting parameters are listed in Table\,\ref{tab:2}. The fitting residuals for $G_{\rm BP}-g, G_{\rm RP}-r, G_{\rm RP}-i, G_{\rm RP}-z, G_{\rm RP}-y$ are 0.008, 0.008, 0.006, 0.009, and 0.012\,mag, respectively. These results suggest that the PS1 magnitudes of individual stars can be predicted with a precision of 1\% or better using the SCR$'$ method, which is similar to the SCR method in \citet{2022AJ....163..185X}. Figure\,\ref{Fig:intrinsic_colors_fitting} demonstrates that there are no dependencies on the \bprpo, \feh, magnitude of PS1, and $E(B-V)_{\rm SFD}$ of the fitting residuals.

Finally, the spatial variations of magnitude offsets for all bands, $\Delta {\bf m}_{\rm SCR'}(\rm R.A., decl.)$, can be obtained from Equations\,(\ref{e2}) and (\ref{e3}).
\begin{eqnarray}
    \begin{aligned}
    {\bf m^{\rm mod}_{\rm SCR'}}=~&{\bf G}_{\rm BP,RP}-{\bf R'} \times E(B-V)_{\rm SFD} - \\
    &{\bf C_{\rm 0}^{\rm mod}}((G_{\rm BP}-G_{\rm RP})_{\rm 0},~\rm [Fe/H])~, \label{e2}
    \end{aligned}
\end{eqnarray}
where, $\bf m^{\rm mod}_{\rm SCR'}$ represents the magnitude predicted by the SCR$'$ method for the PS1 five bands.
    
\begin{eqnarray}
    \Delta {\bf m}_{\rm SCR'}(\rm R.A., decl.)={\bf m^{\rm mod}_{\rm SCR'}}-{\bf m_{\rm PS1}}~,  \label{e3}
\end{eqnarray}    

\begin{figure}[htbp] \centering
\resizebox{\hsize}{!}{\includegraphics{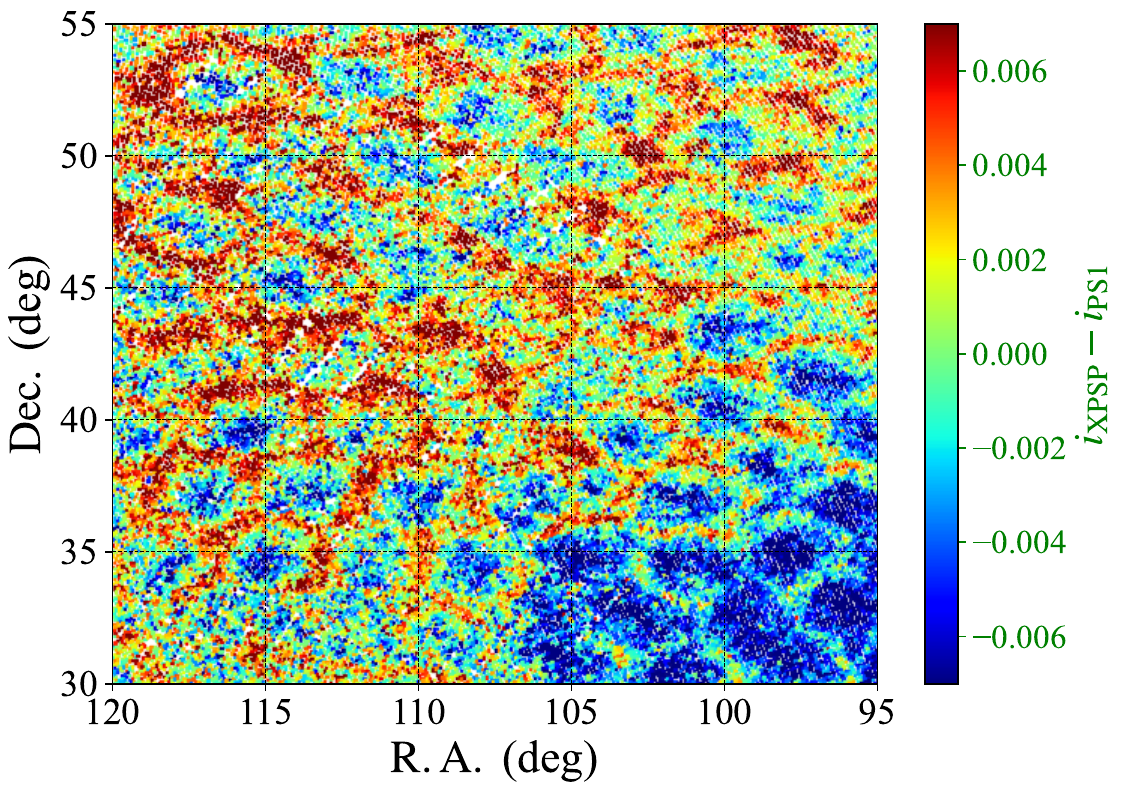}}
\caption{{\small A zoom in plot for a small region in Figure\,\ref{Fig:gspc} in the $i$ band. A color bar is overplotted to the right. Note that the region is the same as Figure\,8 of \citet{2022AJ....163..185X}, but for the XPSP results.}}\label{Fig:ss}
\end{figure}

\begin{figure*}[ht!] \centering
\resizebox{\hsize}{!}{\includegraphics{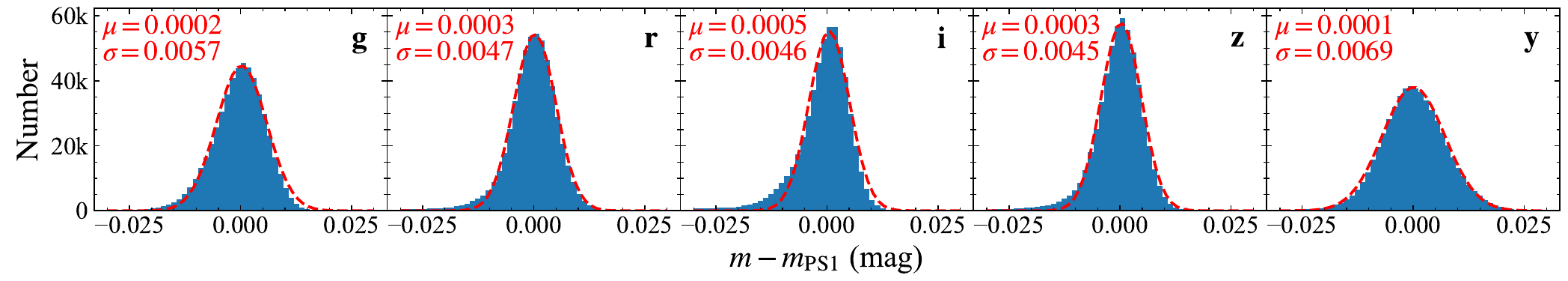}}
\caption{{\small Histogram distributions of derived magnitude offsets from the XPSP method after $14'\times14'$ binning. From left to right are for the $g$, $r$, $i$, $z$, and $y$ bands. The red dotted curves are Gaussian fitting results, with mean and sigma values labeled in the top left corners.}}
\label{Fig:sig}
\end{figure*}

\begin{figure*}[ht!] \centering
\includegraphics[width=13.cm]{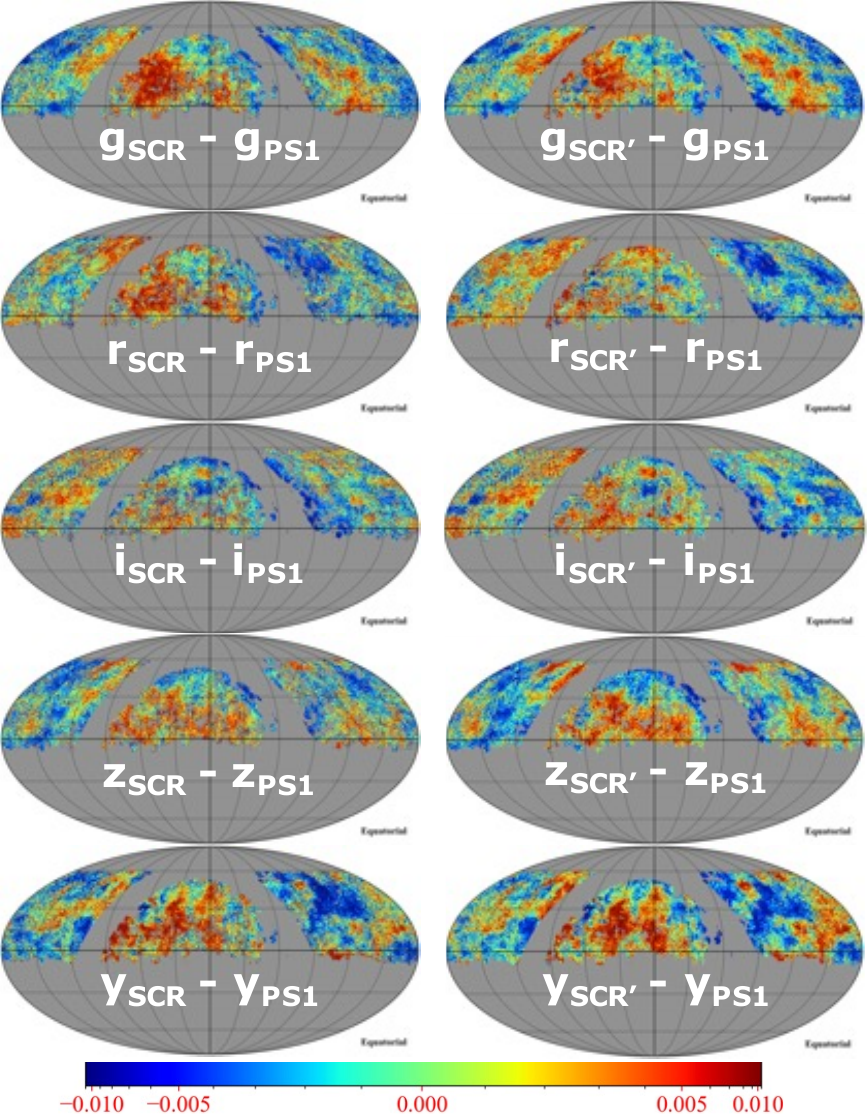}
\caption{{\small Spatial variations of the magnitude offsets of the SCR (left panel) and SCR$'$ (right panel) methods for the common stars in the $g$, $r$, $i$, $z$ and $y$ bands in the Equatorial coordinate system. Color bars are overplotted to the bottom. HEALPix index is level 8.}}
\label{Fig:scr_newscr}
\end{figure*}

\begin{figure*}[ht!] \centering
\includegraphics[width=13.cm]{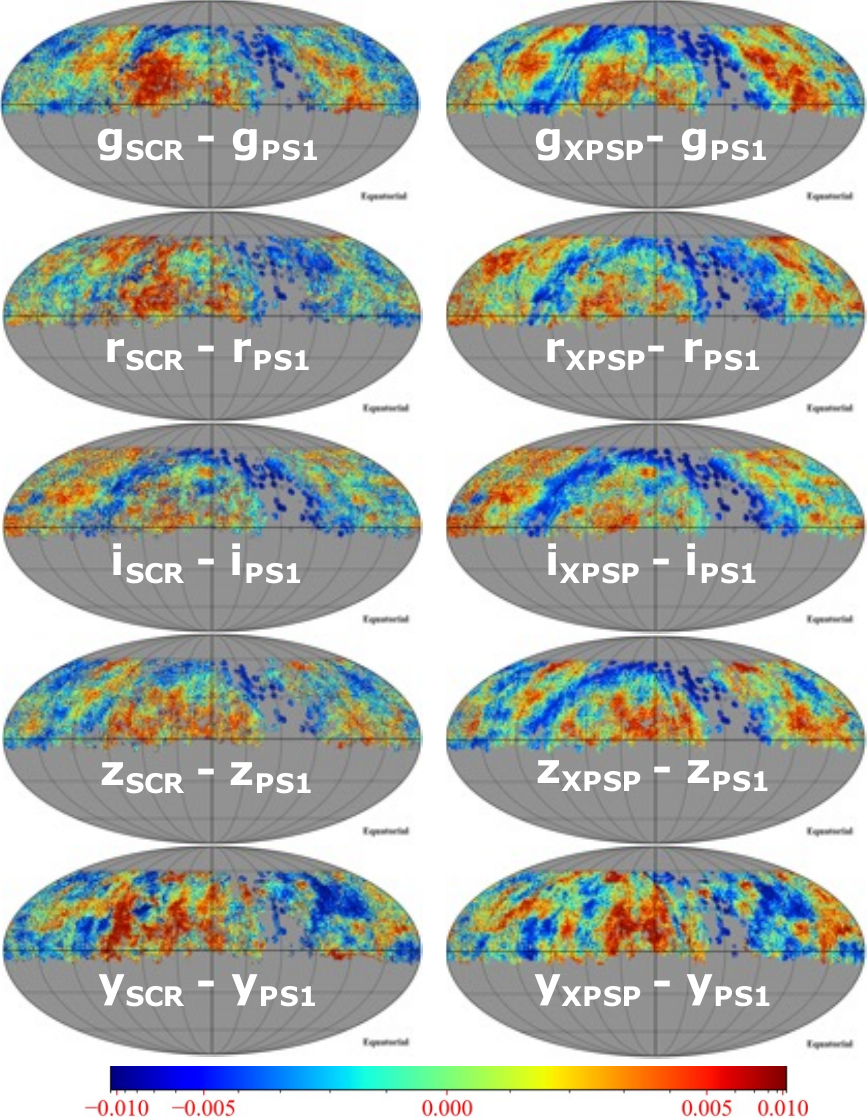}
\caption{{\small Same as Figure\,\ref{Fig:scr_newscr}, but for the SCR (left panel) and XPSP (right panel) methods.}}
\label{Fig:scr_gspc}
\end{figure*}

\begin{figure*}[ht!] \centering
\includegraphics[width=13cm]{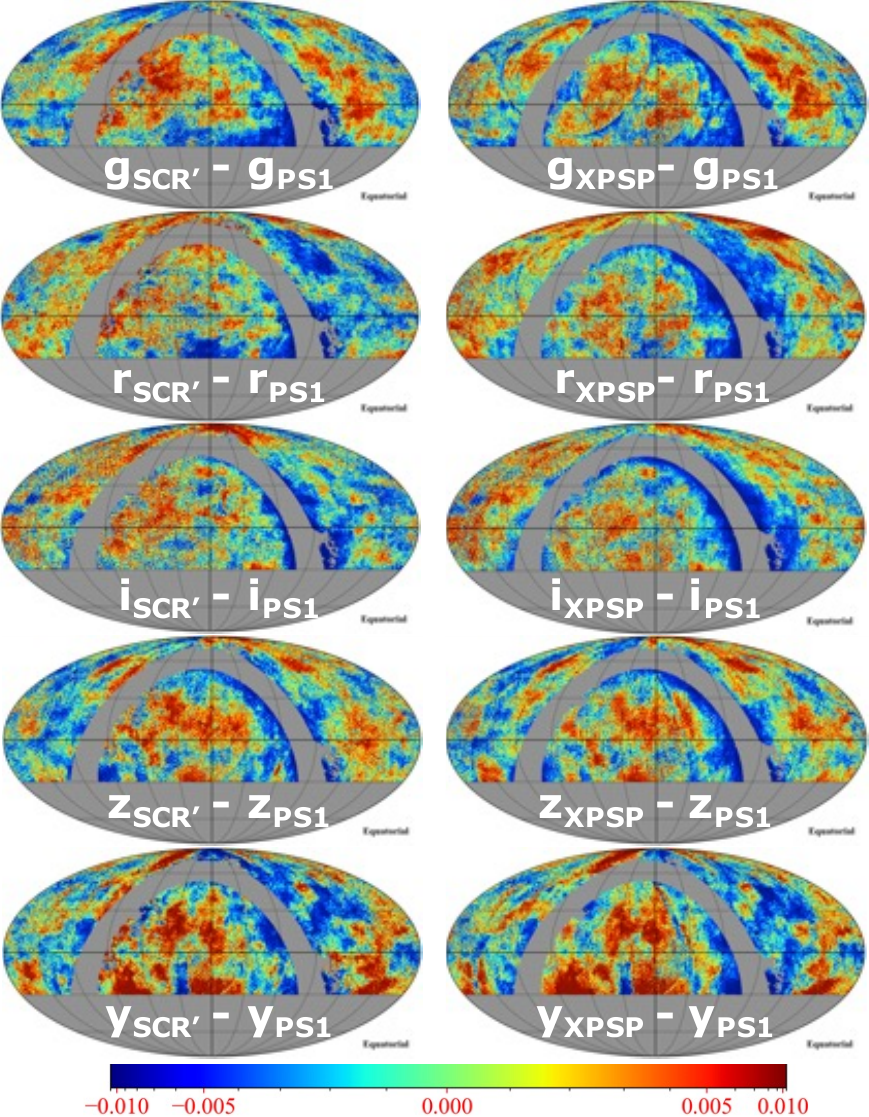}
\caption{{\small Same as Figure\,\ref{Fig:scr_newscr}, but for the SCR$'$ (left panel) and XPSP (right panel) methods.}}
\label{Fig:newscr_gspc}
\end{figure*}

\begin{figure*}[htbp]
  \centering
\resizebox{\hsize}{!}{\includegraphics{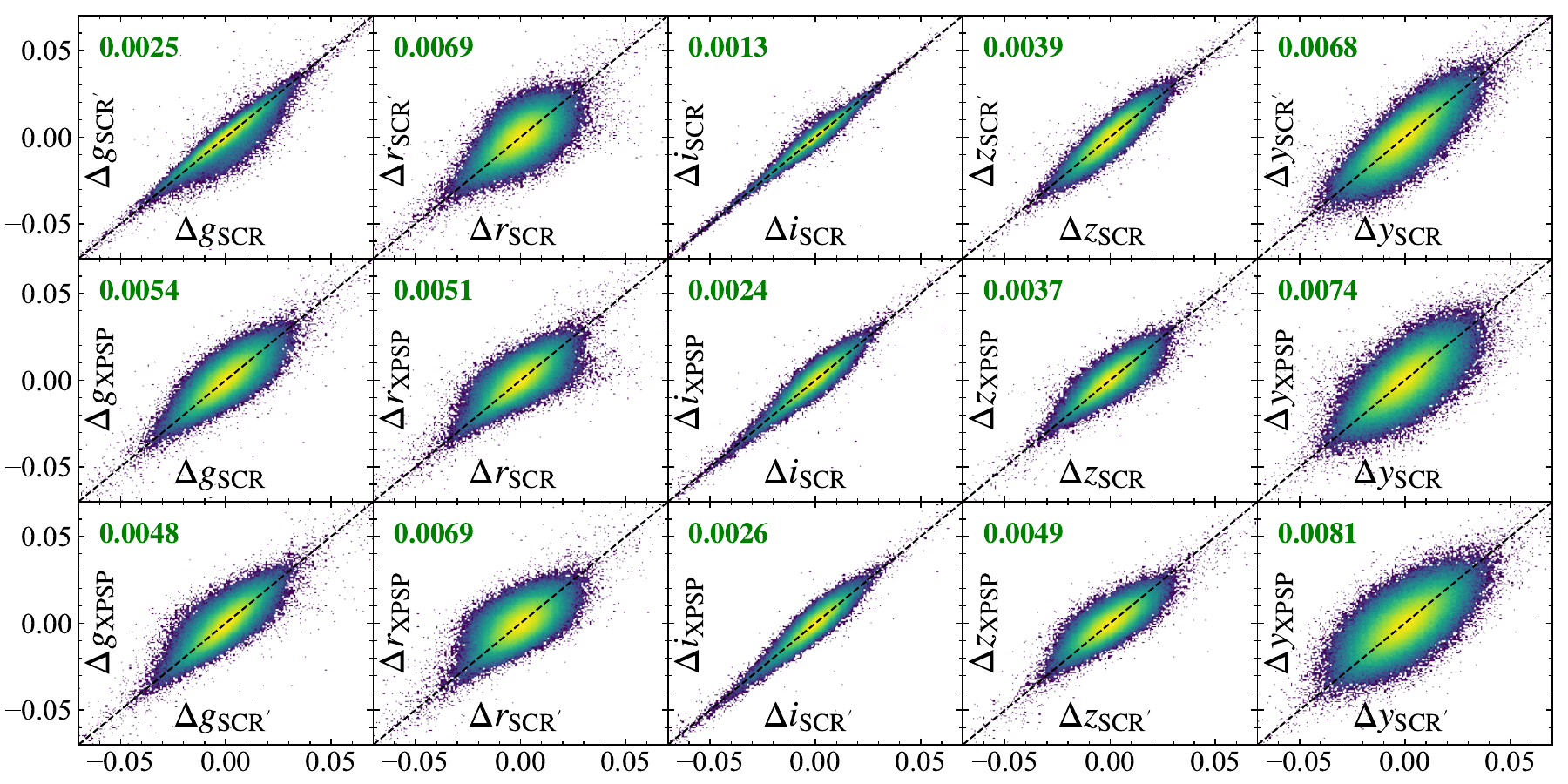}}
\vspace{-3mm}
\caption{{\small The comparison of individual stars for each two methods based on the common stars of three method samples in the $g$, $r$, $i$, $z$ and $y$ bands. For each panel, the standard deviation is marked in the top left corner, and the black dashed line denotes $y=x$. The color in each panel indicates number density of stars. Top: $\Delta m_{\rm SCR'}$ vs. $\Delta m_{\rm SCR}$. Middle: $\Delta m_{\rm XPSP}$ vs. $\Delta m_{\rm SCR}$. Bottom: $\Delta m_{\rm SCR'}$ vs. $\Delta m_{\rm XPSP}$.}}\label{Fig:vsall}
\end{figure*}

\begin{figure*}[htbp]
  \centering
\resizebox{\hsize}{!}{\includegraphics{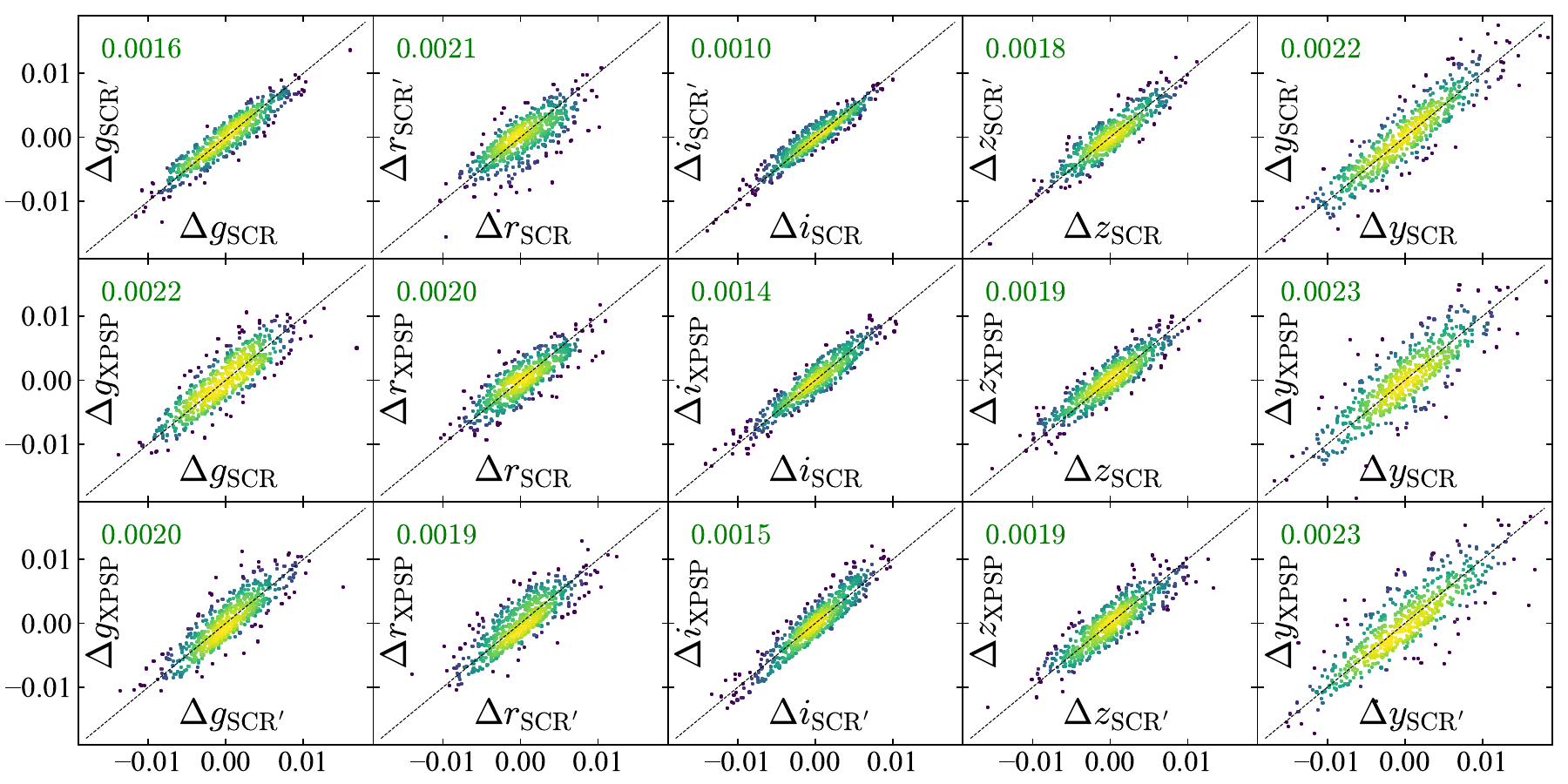}}
\caption{{\small Same as Figure\,\ref{Fig:vsall}, but for the comparison with a restriction of the star numbers of box is more than 20 for SCR, 30 for SCR$'$, and 50 for XPSP methods, after $14'\times14'$ binning.}}\label{Fig:vs}
\end{figure*}

\begin{figure*}[ht!] \centering
\resizebox{\hsize}{!}{\includegraphics{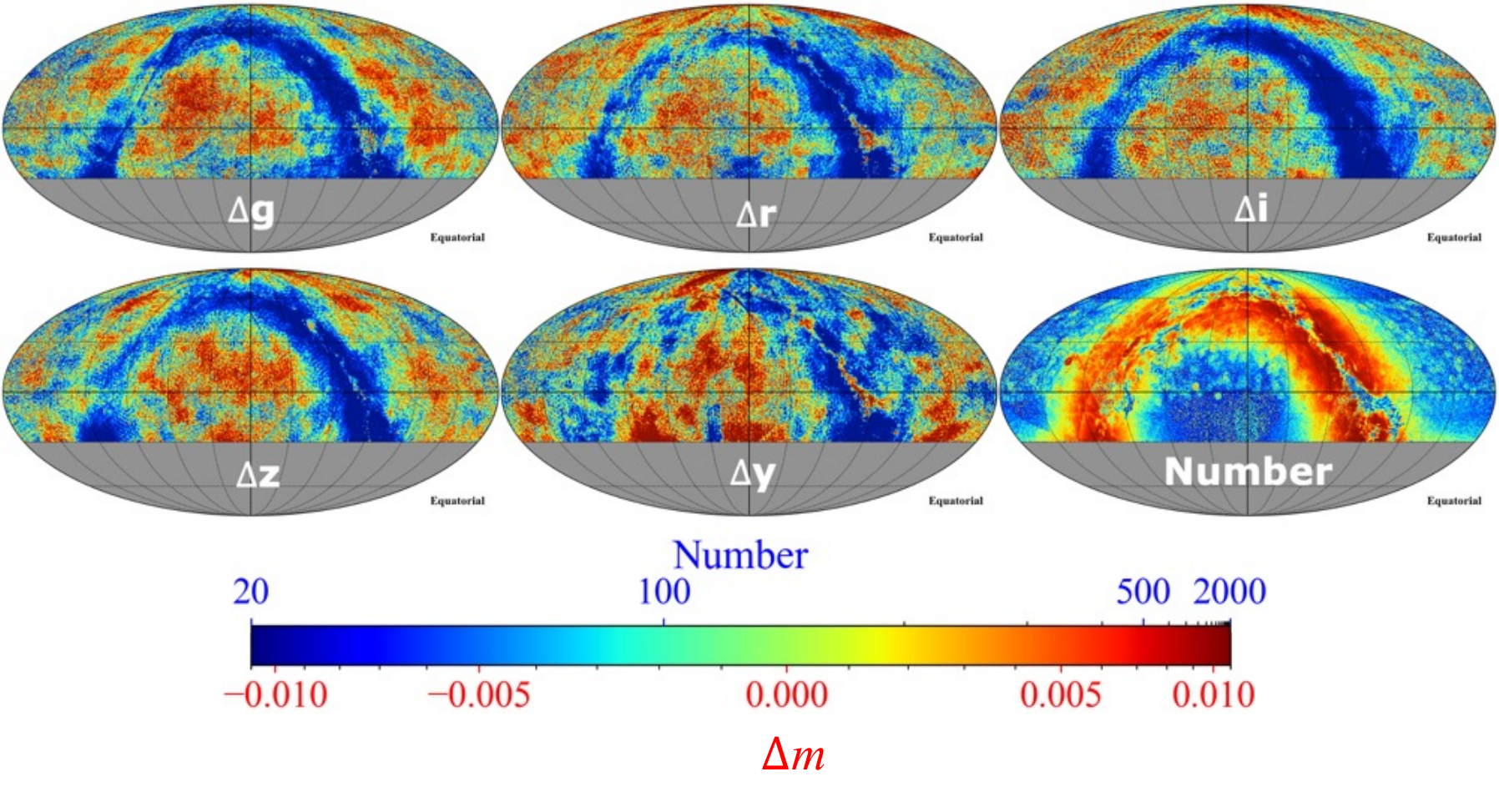}}
\caption{{\small Same as Figure\,\ref{Fig:newscr} but after combining.}}
\label{Fig:corr}
\end{figure*}

\begin{figure*}[ht!] \centering
\resizebox{\hsize}{!}{\includegraphics{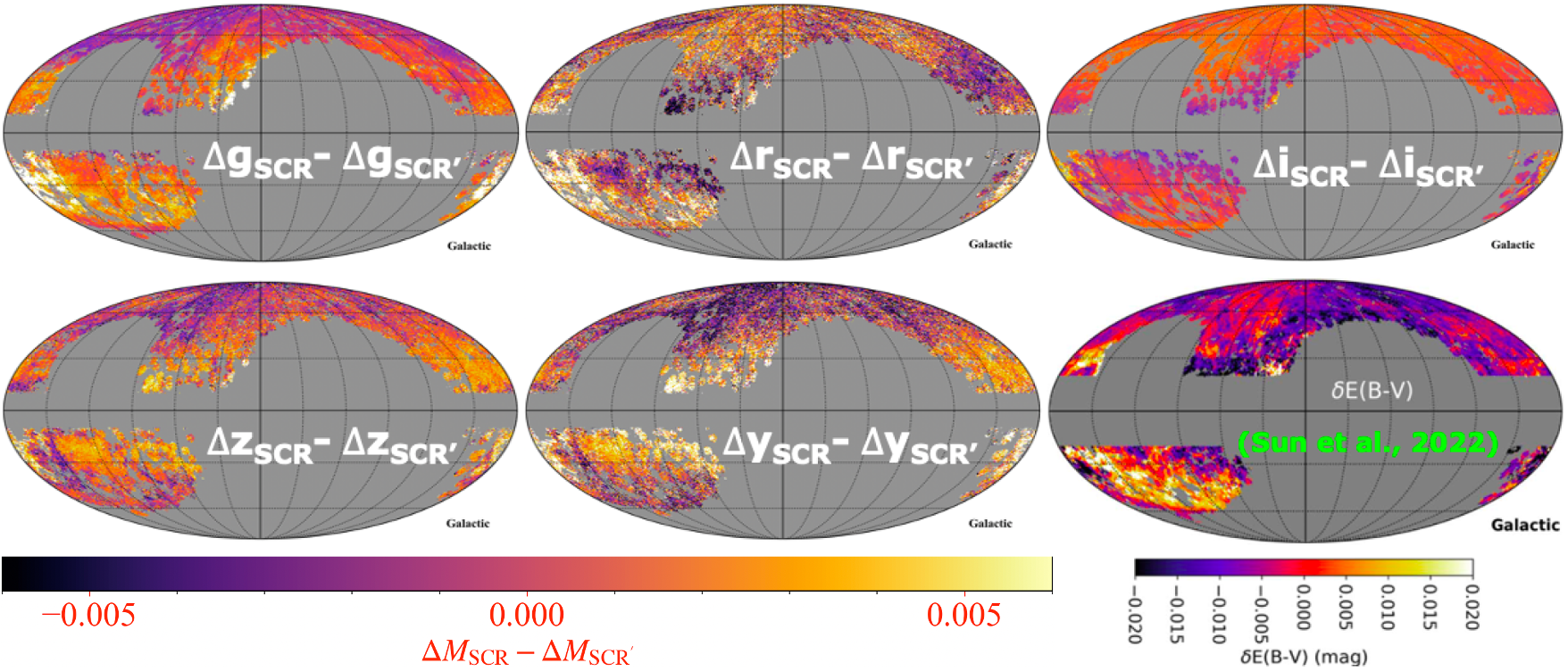}}
\caption{{\small Spatial variations of the difference between the SCR and the SCR$'$ magnitude offsets for the common stars of SCR samples and SCR$'$ samples in the $g$, $r$, $i$, $z$, and $y$ bands in the Galactic coordinate system. All-sky distributions of the spatially-dependent correction from \citet{2022ApJS..260...17S} are shown in bottom right corner panel. Color bars are overplotted to the bottom, respectively.}}
\label{Fig:1minus2}
\end{figure*}

\begin{figure*}[ht!] \centering
\resizebox{\hsize}{!}{\includegraphics{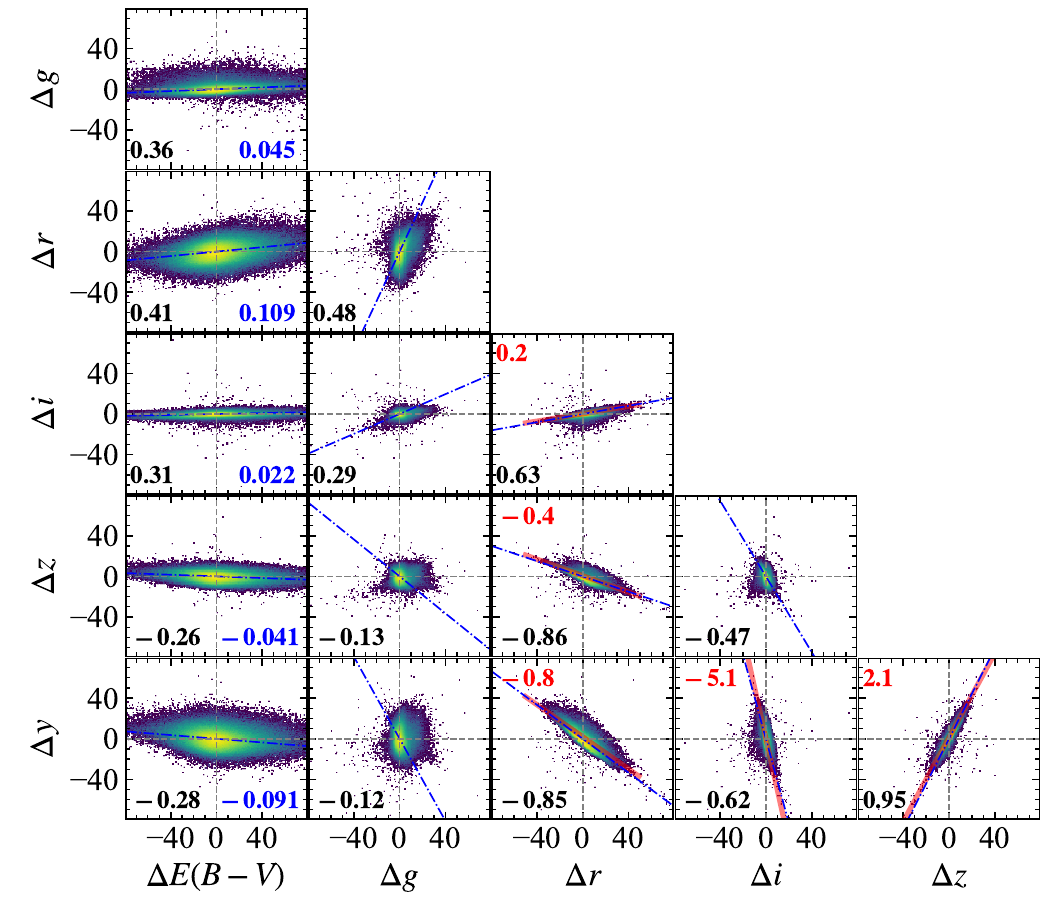}}
\caption{{\small The left panels plot
the offset differences between the SCR and SCR$'$ methods against the spatially dependent errors of $E(B-V)_{\rm SFD}$ in the $grizy$ bands. The other panels compare the differences between the SCR and SCR$'$ magnitude offsets for each two bands. Only boxes with over five stars are plotted.  The color in each panel indicates number
density of stars. The red lines represent the linear fitting results. While the blue lines represent the theoretical results. 
The blue and red lines' slopes are indicated in blue and red respectively, and the correlation coefficients are marked in black.
Here, $\Delta m$ represents $\Delta m_{\rm SCR}-\Delta m_{\rm SCR'}$, $m=\{g,r,i,z,y\}$.}}
\label{Fig:newscr_scr_corr}
\end{figure*}

\begin{figure*}[ht!] \centering
\includegraphics[width=18cm]{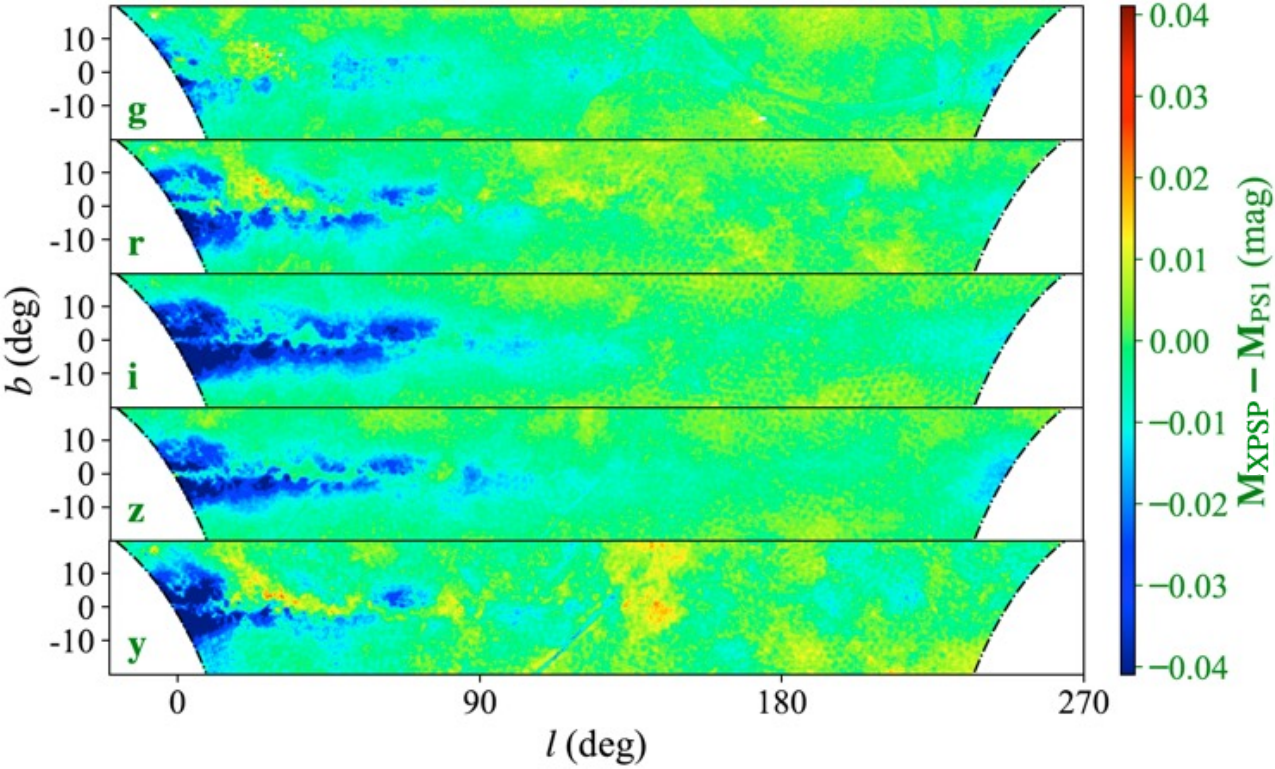}
\caption{{\small Spatial distribution of the XPSP magnitude offsets for stars of $|b|<20^{\circ}$. From top to bottom are for the \grizy~bands, respectively. The color bar is overplotted to the right. The black dotted lines indicate $\rm decl.$ of $-30^{\circ}$.}}
\label{Fig:gspc_r}
\end{figure*}

\begin{figure*}[ht!] \centering
\resizebox{\hsize}{!}{\includegraphics{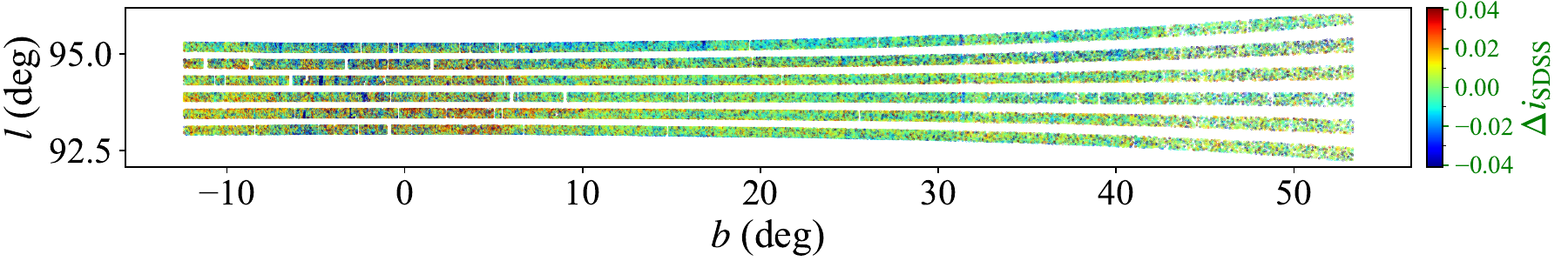}}
\caption{{\small Spatial distribution of the differences between the observed SDSS magnitudes with the synthetic magnitudes of stripe 4682 in the $i$ band. The color bar is overplotted to the right.}}
\label{Fig:tst_sdss}
\end{figure*}

\begin{figure*}[ht!] \centering
\includegraphics[width=18cm]{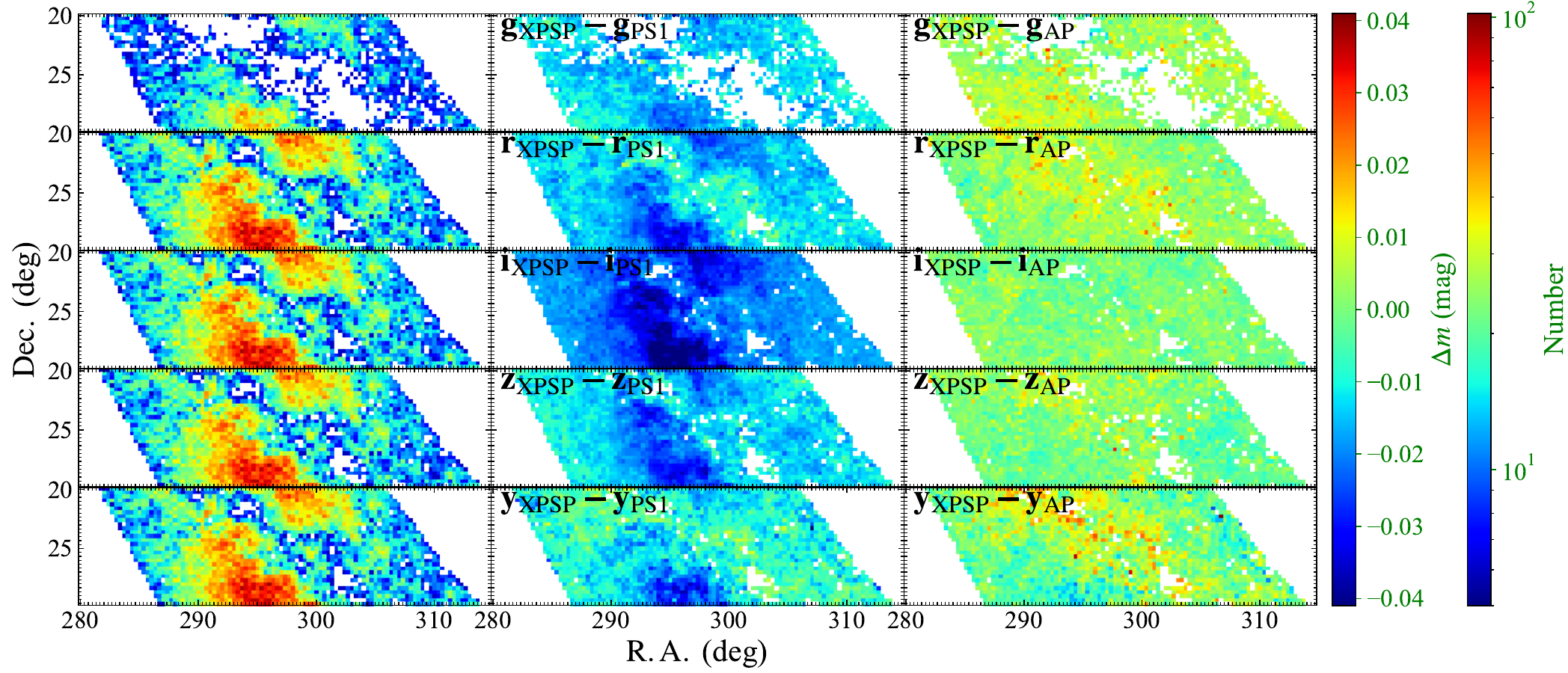}
\caption{{\small Left panels: spatial distribution of the stars. Middle panels: differences between the XPSP standardised magnitudes with the PS1 magnitudes. Right panels: differences between the XPSP standardised magnitudes with the aperture magnitudes of PS1. From top to bottom are for the \grizy~bands, respectively. Color bars are overplotted to the right.}}
\label{Fig:ap_psf}
\end{figure*}

\subsection{The XPSP method} 
\label{sec:m3}
To derive standardized photometry ${\bf m^{\rm mod}_{\rm XPSP}}$ across the entire PS1 survey area, we used the Gaia DR3 XP spectra for approximately 220 million sources as inputs in the $\texttt{GaiaXPy}$ V1.1.4\footnote{\url{https://doi.org/10.5281/zenodo.6674521}} and set \texttt{photometric}\_\texttt{system=PanSTARRS1}\_\texttt{Std}.
For calibration stars, we impose the following constraints: $\texttt{14\,<\,{g, r, i, z} < 17}$, $\texttt{13\,<\,{y}\,<\,17}$, and $\texttt{error (g, r, i, z, y)\,<\,0.02}$\,mag. As a result, 42,442,583, 69,695,226, 93,145,163, 99,742,347, and 105,929,636 stars are selected in the \grizy~bands, respectively. Next, we obtain the spatial variations of magnitude offsets for all bands $\Delta {\bf m}_{\rm XPSP}(\rm R.A., decl.)$ using Equation\,(\ref{e4}):
\begin{eqnarray}
    \Delta {\bf m}_{\rm XPSP}(\rm R.A., decl.)={\bf m^{\rm mod}_{\rm XPSP}}-{\bf m_{\rm PS1}}~.  \label{e4}
\end{eqnarray}

\section{Results} \label{sec:res}
The spatial variations of magnitude offsets obtained by the SCR$'$ and XPSP methods are shown in Figure\,\ref{Fig:newscr} and Figure\,\ref{Fig:gspc}, respectively, after binning with a $14'\times14'$ (HEALPix index level 8) window. Both large- and small-scale spatial patterns are clearly visible and are in excellent agreement with \citet{2022AJ....163..185X}. A zoomed-in plot of the small-scale pattern is shown in Figure\,\ref{Fig:ss}, which is consistent with \citet{2022AJ....163..185X}. 
The small-scale spatial patterns exhibit a typical size of 3 degrees similar to the 3.3 degree FOV of PS1, primarily arise from flat correction errors in PS1 rather than LAMOST, which has a FOV of 5 degrees. The atmospheric parameters (e.g., temperature) derived from LAMOST do not exhibit plate-to-plate systematic errors. These same small-scalle patterns in PS1 are also observed in Figure 10 of \cite{2020ApJS..251....6M}, as mentioned in \cite{2022AJ....163..185X}.
Note that long ring-like patterns are found in Figure\,\ref{Fig:gspc}, particularly in the $g$ and $y$ bands. These patterns come from systematic errors in the Gaia DR3 XP spectra. The spatial structure of minimal systematic errors induced by Gaia aligns with the scanning direction, presenting a clear distinction from the spatial errors observed in PS1.

Based on the results of the XPSP method, we also evaluate the PS1 photometric calibration precisions. Figure\,\ref{Fig:sig} shows the histogram distributions of the position-dependent magnitude offset after $14'\times14'$ binning. The position-dependent magnitude offsets roughly follow Gaussian distributions, with sigma values implying PS1 photometric calibration precisions of 5.7\,mmag, 4.7\,mmag, 4.6\,mmag, 4.5\,mmag, and 6.9\,mmag for the $g$, $r$, $i$, $z$, and $y$ bands, respectively, when averaged over $14'\times14'$ regions.

To test the consistency among the SCR, the SCR$'$, and the XPSP methods, we provide a qualitative comparison of magnitude offsets of common sources among each two methods in Figures\,\ref{Fig:scr_newscr}, \ref{Fig:scr_gspc}, and \ref{Fig:newscr_gspc}.  
The results reveal that the magnitude offsets obtained by the three methods are consistent, with best consistency observed in the $i$ band. 
A quantitative comparison for individual stars is illustrated in Figure\,\ref{Fig:vsall}. All sources in each panel are distributed close to the line of $y=x$. 
The standard deviations of the difference between magnitude offsets for each two methods are around 3--5, 5--7, 1--3, 4--5, and 7--8\,mmag in the $g$, $r$, $i$, $z$, and $y$ bands, respectively.
After $14'\times14'$ binning,  Figure\,\ref{Fig:vs} shows that the standard deviations  were significantly reduced, with values ranging from 1--2\,mmag.

To correct the spatially dependent systematic errors of PS1, we combine the magnitude offsets of individual calibration stars from the three methods with weights of 4 (the SCR method), 2 (the SCR$'$ method), and 1 (the XPSP method), and perform adaptive median smoothing with an initial box size of $14'\times14'$ (HEALPix index level 8). If the number of stars within a box is less than 20, we increase the HEALPix index until it reaches level 5 (1.8$^\circ$). Note that only less than 1 percent of pixels have a smoothing radius greater than $14'\times14'$. 

The results of the five bands after smoothing are plotted in Figure\,\ref{Fig:corr} and can be used to correct calibration errors in the whole PS1 data. The maps are publicly available\footnote{\url{https://doi.org/10.12149/101283}}. The corrected magnitudes $m^{\rm corr}$ can be computed as
\begin{eqnarray}
  \begin{aligned}
  {\bf m}^{\rm corr}=~&{\bf m}_{\rm PS1}+\Delta {\bf m}(\rm R.A., \rm decl.)~, 
  \end{aligned}
\end{eqnarray}
where ${\bf m}_{\rm PS1}$ is the magnitude after magnitude-dependent errors corrected by \citet{2022AJ....163..185X}, and $\Delta {\bf m}(\rm R.A., \rm decl.)$ is the spatial-dependent magnitude offset.

\section{Discussion} \label{sec:discussion}
\subsection{Comparison of the SCR and SCR$'$ methods} \label{sec:s1}
Detailed comparisons between the SCR and SCR$'$ magnitude offsets are displayed in Figure\,\ref{Fig:1minus2}. It can be seen that the scatter in the $r$ and $y$ bands is larger than in the other bands, and correlations between the different bands have emerged, for example, a negative correlation between the $r$ and $y$ bands, and a positive correlation between the $z$ and $y$ bands.

The SCR$'$ method used the dust reddening map of \citet{1998ApJ...500..525S} for reddening correction. \citet{2022ApJS..260...17S} have carried out a systematic and comprehensive validation of the reddening map of \citet{1998ApJ...500..525S} against about 2 million stars based on spectroscopic data from LAMOST and photometric data from Gaia using the star-pair method \citep{2013MNRAS.430.2188Y,2020ApJ...905L..20R}. They found spatially dependent errors of up to 10 to 20 mmag in $E(B-V)_{\rm SFD}$, as shown in the bottom right panel of Figure\,\ref{Fig:1minus2}.

To investigate whether the magnitude offsets derived from the SCR$'$ method are affected by possible systematic errors in $E(B-V)_{\rm SFD}$, we calculate and plot the correlations between the spatially dependent errors of $E(B-V)_{\rm SFD}$ and the difference between the SCR and SCR$'$ magnitude offsets.
Only boxes with star numbers larger than 5 in all the five bands are used. These correlations are shown in the first column of Figure\,\ref{Fig:newscr_scr_corr}. Furthermore, the correlations between the difference in the SCR and SCR$'$ magnitude offsets for each two bands with the same restriction are also displayed in Figure\,\ref{Fig:newscr_scr_corr}.  We perform linear regressions on the magnitude offsets between two bands that exhibit a significant correlation. The slopes are shown in Figure\,\ref{Fig:newscr_scr_corr}.

The impact of the spatially dependent systemic errors of $E(B-V)_{\rm SFD}$ on the magnitude offsets in the SCR$'$ method is solely determined by the reddening values and the intrinsic color fitting polynomial.
Theoretical calculations were conducted to quantify the incremental alterations in both the reddening values and intrinsic color of the five colors, resulting from a 0.01\,mag rise in $E(B-V)_{\rm SFD}$.
By adding the increments of both the reddening value and intrinsic color, one can determine the resulting increase in magnitude ($\delta {\bm m}$) for the $grizy$ bands in SCR$'$, which are 0.45, 1.09, 0.22, $-$0.41, and $-$0.91\,mmag, respectively.
The theoretical lines with a slope of $\delta {\bm m}/0.01$ are over-plotted in the left panels of Figure\,\ref{Fig:newscr_scr_corr}.
We can see that the typical magnitude offsets in Figure\,\ref{Fig:newscr_scr_corr} exhibit excellent consistency with theoretical calculations. For example, the theoretical ratio of $\Delta z$ versus $\Delta y$ is $\frac{\Delta z{\rm SCR}-\Delta z_{\rm SCR'}}{\Delta y_{\rm SCR}-\Delta y_{\rm SCR'}}=\frac{-0.91\times \Delta E(B-V){\rm SFD}}{-0.41\times \Delta E(B-V){\rm SFD}}=2.2$, which is consistent with the slope of 2.1 obtained from the linear regression.

The spatially dependent errors of $E(B-V)_{\rm SFD}$ are the primary cause of the difference between the SCR and SCR$'$ magnitude offsets in all bands, with a particularly noticeable impact in the $r$ and $y$ bands. The correlations between the difference in magnitude offsets for each pair of bands shown in Figure\,\ref{Fig:newscr_scr_corr} can also be attributed to these errors. Specifically, we have estimated that for the SCR$'$ method, the magnitude offset errors in the \grizy~bands are approximately 1\,mmag, 2\,mmag, 0.4\,mmag, 1\,mmag, and 2\,mmag, respectively, when the typical value of $E(B-V)_{\rm SFD}$ error is 0.02\,mag. This explains why the scatter in the $r$ and $y$ bands is relatively larger than that in other bands as shown in the top panels of Figure\,\ref{Fig:vsall} and \ref{Fig:vs}. Conversely, the $i$ band exhibits the smallest scatter, as the $G_{\rm RP}-i$ color is most insensitive to errors of $E(B-V)_{\rm SFD}$.

We note that the photometric metallicities presented by \citet{2022ApJS..258...44X} suffer moderate spatially dependent systematic errors, up to 0.2\,dex. However, the impact of such errors on the PS1 passbands is small. In the case of blue filters that are sensitive to metallicity, these errors may cause moderate spatially dependent systematic errors when using the SCR$'$ method for calibration, and therefore cannot be disregarded.

\subsection{The magnitude offsets in the Galactic plane}

In Figure\,\ref{Fig:gspc}, we can clearly see that there are bad matches (up to about 0.04 mag) between the PS1 magnitudes $m_{\rm PS1}$ and the standardized magnitudes $m_{\rm XPSP}$ in the Galactic plane, especially for the $i$ band. The discrepancies are more clearly displayed in Figure\,\ref{Fig:gspc_r}. Similar result was also reported in \citet{2020ApJS..251....6M} by comparing the Gaia DR1 photometry in the $G$ band with the transformed Gaia $G$ magnitudes from PS1. They attributed this mismatch to the transformation process. 

To investigate the possible causes, we have performed several checks. First, we compare the observed SDSS  magnitudes with the synthetic SDSS XPSP magnitudes for stars in stripe 4682 as an example. This stripe crossed the inner Galactic disk. Figure\,\ref{Fig:tst_sdss} shows the spatial distribution of the differences in the $i$ band. No significant PS1-likely offsets are found in the low Galactic latitude region. This suggests that Gaia XP spectra are probably not the cause for the bad match. It is further confirmed by the very weak dependence of the magnitude differences on either \ebprp ~or \bprp~for LAMOST stars in the low Galactic latitude region.  

We finally find that the bad match in the Galactic plane is probably caused by the systematic errors of the PS1 PSF magnitudes in crowded fields. We randomly selected approximately 50,000 stars located in the Galactic plane, restricted to $\rm 280<R.A.<315$ and $\rm 20<decl.<30$ from PS1 DR1 for $i$ band. Of these sources, 22,280, 49,409, 46,380, and 48,891 have $grzy$-band magnitudes, respectively. The middle and right panels of Figure\,\ref{Fig:ap_psf} show the comparison between their XPSP magnitudes with PSF magnitudes and aperture-based magnitudes, respectively. The XPSP magnitudes exhibit good agreement with the aperture-based magnitudes in each band. It suggests that systematic errors in the PSF magnitudes of PS1 account for the bad match. The mismatch is strongly correlated with the density of stars, as demonstrated in Figure\,\ref{Fig:ap_psf}.

\section{Conclusions} \label{sec:conclusion}

In this work, we have re-calibrated the PS1 photometry by correcting for position-dependent systematic errors using the SCR method based on the corrected Gaia EDR3 and the spectroscopic data from LAMOST DR7, the SCR$'$ method based on the photometric metallicities of \citet{2022ApJS..258...44X}, and the XPSP method based on the Gaia DR3 XP spectra. Using 50--100 million stars as standards, the significant large-scale and small-scale spatial variation of magnitude offsets caused by calibration errors are revealed in the whole PS1 footprint. The calibration errors when averaged over 14$'$ regions are approximately 5.7, 4.7, 4.6, 4.5, and 6.9\,mmag  for the $grizy$ filters, respectively. 
The results of the three methods are consistent with each other within 1--2\,mmag or better for
all the filters. 

We have identified systematic errors up to 0.04\,mag or larger in the Galactic plane, with were mostly likely attributed to systematic errors in the PSF magnitudes of PS1. 
We also discussed the differences between the SCR and SCR$'$ methods, with the main cause being the position-dependent systematic errors of $E(B-V)_{\rm SFD}$. 

We have provided two-dimensional maps and a python package to correct for position-dependent magnitude offsets. These maps, together with the magnitude-dependent correction of \cite{2022AJ....163..185X}, can serve as a better reference to calibrate other surveys.

\begin{acknowledgments}
This work is supported by the National Natural Science Foundation of China through the project NSFC 12222301, 12173007 and 11603002,
the National Key Basic R\&D Program of China via 2019YFA0405503 and Beijing Normal University grant No. 310232102. 
We acknowledge the science research grants from the China Manned Space Project with NO. CMS-CSST-2021-A08 and CMS-CSST-2021-A09.

This work has made use of data from the European Space Agency (ESA) mission Gaia (\url{https://www.cosmos.esa.int/gaia}), processed by the Gaia Data Processing and Analysis Consortium (DPAC, \url{https:// www.cosmos.esa.int/web/gaia/dpac/ consortium}). Funding for the DPAC has been provided by national institutions, in particular the institutions participating in the Gaia Multilateral Agreement. 
Guoshoujing Telescope (the Large Sky Area Multi-Object Fiber Spectroscopic Telescope LAMOST) is a National Major Scientific Project built by the Chinese Academy of Sciences. Funding for the project has been provided by the National Development and Reform Commission. LAMOST is operated and managed by the National Astronomical Observatories, Chinese Academy of Sciences.

The Pan-STARRS1 Surveys (PS1) and the PS1 public science archive have been made possible through contributions by the Institute for Astronomy, the University of Hawaii, the Pan-STARRS Project Office, the Max-Planck Society and its participating institutes, the Max Planck Institute for Astronomy, Heidelberg and the Max Planck Institute for Extraterrestrial Physics, Garching, The Johns Hopkins University, Durham University, the University of Edinburgh, the Queen's University Belfast, the Harvard-Smithsonian Center for Astrophysics, the Las Cumbres Observatory Global Telescope Network Incorporated, the National Central University of Taiwan, the Space Telescope Science Institute, the National Aeronautics and Space Administration under Grant No. NNX08AR22G issued through the Planetary Science Division of the NASA Science Mission Directorate, the National Science Foundation Grant No. AST–1238877, the University of Maryland, Eotvos Lorand University (ELTE), the Los Alamos National Laboratory, and the Gordon and Betty Moore Foundation.

\end{acknowledgments}

\clearpage
\appendix
\setcounter{table}{0}   
\setcounter{figure}{0}
\renewcommand{\thetable}{A\arabic{table}}
\renewcommand{\thefigure}{A\arabic{figure}}
\section {Estimate the temperature from the intrinsic color $(G_{\rm BP}-G_{\rm RP})_{\rm 0}$ and metallicity}
\begin{figure*}[ht!] \centering
\includegraphics[width=15.cm]{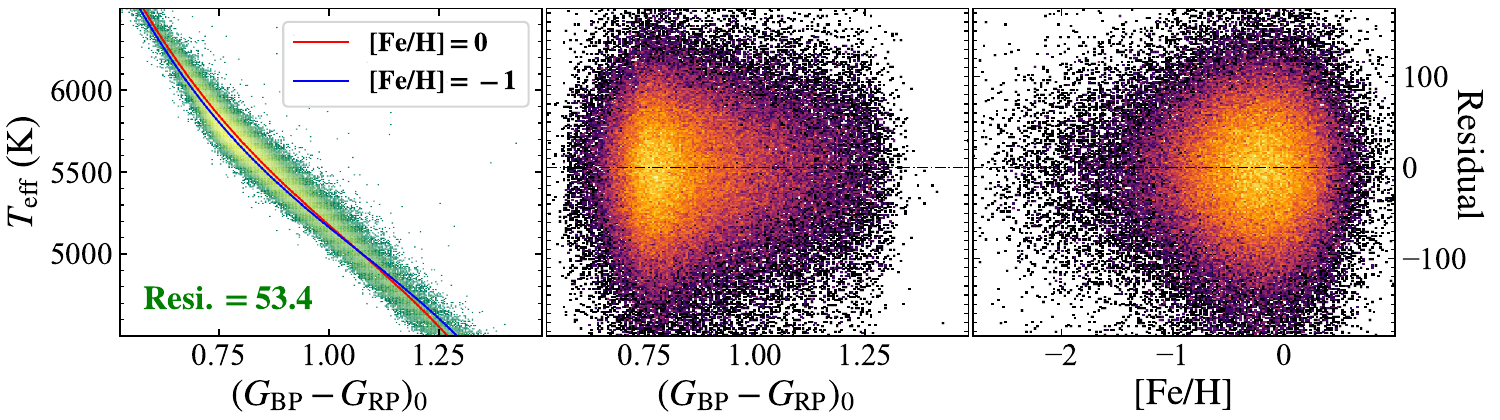}
\caption{{\small Two-dimensional third-order polynomial fitting (with ten free parameters) of temperature as functions of \bprpo~and \feh~for around 110,000 stars. The left panel shows the fitting results, the red and blue curves represent results for $\rm [Fe/H]$ = 0 and $-$1, respectively, and the fitting residuals are labeled. The middle, right panel plot residuals against \bprpo, \feh~, respectively. The black dotted lines denote residuals of 0.}}
\label{Fig:teff_color0}
\end{figure*}
To estimate the temperatures of calibration stars in the SCR$'$ method, we selected a sample of 110,000 stars by combining calibration stars with LAMOST DR7. A cross-matching radius of 1$''$ was adopted, and a constraint of $E(B-V)<0.2$ was applied to reduce the effect of extinction. We then used a three-order two-dimensional polynomial function of $(G_{\rm BP}-G_{\rm RP})_{\rm 0}$ and metallicity $\rm [Fe/H]$ to fit the temperature $T_{\rm eff}$ of the sample from LAMOST DR7. The resulting fit, shown in Figure,\ref{Fig:teff_color0}, had a fitting residual of 53\,K, indicating a precision of 1\% or better. The relationship between temperature and $(G_{\rm BP}-G_{\rm RP})_{\rm 0}$ and $\rm [Fe/H]$ is described by the equation $T_{\rm eff}^{\rm mod}=-3465.31x^3+1.13y^3-286.24x^2y+16.20xy^2+10771.58x^2+20.50y^2+409.88 y-13512.97\times x-74.37y+11387.71$, where $x$ represents $(G_{\rm BP}-G_{\rm RP})_{\rm 0}$, and $y$ represents $\rm [Fe/H]$. The fitting residuals showed no dependence on $(G{\rm BP}-G_{\rm RP})_{\rm 0}$ and $\rm [Fe/H]$.

\end{document}